\documentclass[apj]{emulateapj}
\usepackage{times}
\usepackage{color}
\usepackage{epsf}
\usepackage{graphicx}
\usepackage{ulem}

\newcommand{\rf}[1]{\ref{fig:#1}}

\def\knew{\kappa_{\rm ln}}
\newcommand{\pderiv}[2]{\frac{\partial#1}{\partial#2}}
\newcommand{\Cl}[2]{C_{#2}(#1)}

\newcommand{\hMpc}{\,h^{-1}{\rm Mpc}}

\def\lcdm{$\Lambda$CDM~}
\def\eg{{\it e.g.}}
\def\ie{{\it i.e.}}

\def\be{\begin{equation}}
\def\ee{\end{equation}}
\def\bea{\begin{eqnarray}}
\def\eea{\end{eqnarray}}

\newcommand{\paone}{Paper I}

\newcommand{\Nb}{{\it N}-body}
\newcommand{\lmax}{l_{\rm max}}
\newcommand{\jcap}{JCAP}
\newcommand{\Om}{\Omega_m}
\newcommand{\Ox}{\Omega_X}
\newcommand{\Ob}{\Omega_b}
\newcommand{\Ok}{\Omega_K}
\newcommand{\Obhh}{\Omega_bh^2}
\newcommand{\Omhh}{\Omega_mh^2}
\newcommand{\Oh}{\Omega_mh^2}
\newcommand{\Ochh}{\Omega_ch^2}
\newcommand{\lnAs}{\ln A_s}

\newcommand{\FoMw}{\rm FoM_{w_0-w_a}}
\newcommand{\iams}{{\rm arcmin}^{-2}}
\newcommand{\dl}{\Delta l}
\def\kamin{\kappa_{\rm min}}
\def\cov{{\rm Cov}}
\shorttitle{}
\shortauthors{}
\begin{document}

\title{Dark Energy from the log-transformed convergence field}
\author{Hee-Jong Seo\altaffilmark{1}, Masanori Sato\altaffilmark{2}, Masahiro Takada\altaffilmark{3}, and Scott Dodelson\altaffilmark{4,5}}
\altaffiltext{1}{Berkeley Center for Cosmological Physics, LBL and Department of Physics, University of California, Berkeley, CA, USA 94720. hee-jongseo@lbl.gov}
\altaffiltext{2}{Department of Physics, Nagoya University, Nagoya 464-8602, Japan}
\altaffiltext{3}{Institute for the Physics and Mathematics of the
Universe (IPMU), University of Tokyo, Chiba 277-8582, Japan}
\altaffiltext{4}{Center for Particle Astrophysics, Fermi National
Accelerator Laboratory, Batavia, IL~~60510}
\altaffiltext{5}{Department of Astronomy \& Astrophysics, The
University of Chicago, Chicago, IL~~60637}
\begin{abstract}
A logarithmic transform of the convergence field improves `the information content', \ie, the overall precision associated 
with the measurement of the amplitude of the convergence power spectrum by improving the covariance matrix properties. The translation of this improvement in the information content to that in cosmological parameters, such as those associated with dark energy, 
requires knowing the sensitivity of the log-transformed field to 
 those cosmological parameters. In this 
paper we use \Nb\ simulations with ray tracing to generate convergence fields at multiple source redshifts as a function of cosmology. The gain in information associated with the log-transformed field does lead to tighter constraints on dark energy parameters, but only if shape noise is neglected. The presence of shape noise quickly diminishes the advantage of the log mapping, more quickly than we would expect based on the information content. With or without shape noise, using a larger pixel size allows for a more efficient log-transformation. 
\end{abstract}


\section{Introduction}
The nature of the dark energy is one of the most intriguing mysteries of the Universe. As a result, various large sky area surveys are being conducted and designed to statistically determine the properties of this energy component with high precision. We of course wish to extract the most information available from the data. Recently, it has been suggested that the two-point statistics of the logarithmically transformed nonlinear density or weak-lensing convergence field may contain more information than the conventional nonlinear fields without the transformation~\citep[e.g.,][]{Neyrinck09,SeoKappa11}. In linear theory (i.e., at high redshift), the convergence (and density) field is Gaussian, which means that the two-point function contains all the information. Due to the structure growth, the field however becomes more nonlinear and non-Gaussian at low redshift. The cosmological information in the two-point function therefore decreases with increasing nonlinearity, the lost information moving to the higher order statistics \citep{Takada:2003ef}. 
  
A logarithmic transform of the nonlinear mass/galaxy density field or the weak-lensing convergence field, which makes the one-point distribution of the field more Gaussian~\citep[e.g.,][]{Coles91,Kayo01,Taruya02}, appears to produce a final field that alleviates this problem by mimicking properties of a Gaussian field~\citep{Neyrinck09,SeoKappa11,Yu11}. The two-point function of the transformed field has a more diagonal covariance matrix~\citep[e.g.,][]{Neyrinck11a} and (therefore) increased information content (\ie, the overall precision associated with the measurement of the amplitude of the convergence/density power spectrum) to a level comparable to the Gaussian field. \citet{SeoKappa11} showed that a Taylor expansion of the logarithmic transformation that includes up to the bispectrum contribution captures most of the improvement on large scales, suggesting that the log-transform draws its extra information from higher order statistics.
 
In these works, the benefits of the log-transform have been studied only when a single parameter -- the amplitude of the power spectrum -- is varied. 
This does not necessarily translate into an improvement in the measurements of other cosmological parameters, such as dark energy parameters. If the power spectrum of the transformed field is less sensitive to, \eg, dark energy (\ie, smaller derivatives with respect to the parameters), or suffers more degeneracies between parameters, the log-transform may not be as effective as expected based on the improved information content. In order to test this, we need to understand the dependence of the log-transformed field as a function of cosmology.
In \citet{SeoKappa11}, 
which we refer to as Paper I hereafter,
 we used a modified log-transform for the weak-lensing convergence field and showed the increased information content after the log transform. Here we extend the previous work and investigate whether or not this improvement propagates to the determination of the dark energy parameters. A first step in this direction was taken in \citet{Joachim11} for the weak lensing field, wherein mapping the required derivatives with respect to cosmological parameters were computed analytically. They showed that the Box-Cox transformation that encloses the logarithmic transformation, when optimized, indeed gives a tighter constraint on $\Om-\sigma_8$. Also, \citet{Neyrinck11c} has recently shown that, using Coyote Universe simulations \citep{Coyote10}, log-transformed density field gives tighter, but unmarginalized, constraints on cosmological parameters.

Here, we adopt a more extensive numerical approach and derive the dependence of the transformed field on the cosmological parameters directly from a very large set of \Nb\ simulations produced for various cosmologies. The total simulated sky is $37000$ square degrees. We produce mock weak-lensing convergence fields at three source redshifts, carry out a Fisher matrix analysis of tomography using the fully nonlinear covariance matrix and derivatives, and derive marginalized errors on 6-8 cosmological parameters, including dark energy parameters (\ie, $\Ox$, $w_0$, and $w_a$). As far as we know, this paper is the first to conduct the full Fisher matrix analysis of weak-lensing power spectrum tomography using the numerical simulations to compute the derivatives and covariance matrices \citep[but also see][for peak statistics tomography]{Yang11}.

The paper is organized as following. In \S~\ref{sec:method}, we explain the details of our \Nb\ simulations and Fisher matrix analysis.
 In \S~\ref{sec:IC}, we revisit the general properties of the log transform, such as the one-point probability distribution function (PDF), the information content, and the 
covariance matrix from the \Nb\ results using three source redshift bins. 
In \S~\ref{sec:Fisher}, we present the numerical derivatives before and 
after the logarithmic mapping and the results of the full Fisher matrix 
analysis, deriving the dark energy figure-of-merit. We compare this result with the Fisher matrix 
obtained from semi-analytic fits to the power spectrum. In \S~\ref{sec:Fshape}, we include shape noise; in \S~\ref{sec:Pixel}, we discuss the effect of the size of pixels; in \S~\ref{sec:Fhalofit}, we discuss the analytic, Gaussian Fisher matrix results in comparison to our \Nb\ results for the fiducial field. Finally, we conclude in \S~\ref{sec:con}. 

\section{Numerical Fisher matrix analysis}
\label{sec:method}
To study the cosmological information from the log-transformed field
numerically, we use a large set of ray-tracing simulations.
The ray-tracing simulations are constructed from 2$\times$200
realizations of \Nb\ simulations with box sizes of 240 and
480$h^{-1}$Mpc on a side, respectively. 
The number of particles in each simulation is 256$^3$.
For the fiducial cosmology, we adopt the concordance \lcdm model: matter 
fraction $\Om = 0.238$, baryon fraction $\Ob = 0.042$, dark energy fraction 
$\Ox = 0.762$ (therefore a flat universe), the equation of state parameters 
$w_0 = -1$ and $w_a=0$, spectral index $n_s = 0.958$, normalization $A_s=2.35\times 10^{-9}$,  
and Hubble parameter $h = 0.732$. 
Note that the fiducial cosmology gives $\sigma_8 = 0.76$ 
(the variance of the 
present-day
density fluctuation in a sphere of radius $8\hMpc$). 
We assume three delta-function like source 
redshifts at $z=0.6$, $1.0$, and $1.5$ for a tomographic study. From the 400 \Nb\ simulations, we generate 1000 realizations of 
$5^\circ \times 5^\circ$ 
lensing convergence
fields (i.e., a total of 25000 square degrees)
 with $2048^2$ pixels
(0.15 arcmin per pixel) at each source
 redshift using ray-tracing 
(a total of 3000 convergence fields). 
Details of the ray-tracing can be found in \citet{Sato09} \citep[see,
also][]{Sato11}.

We resample the convergence fields in $128^2$ pixels 
(2.4 arcmin per pixel) by averaging over nearby 16 by 16 pixels
 as our fiducial case. As will be discussed in \S~\ref{sec:Pixel}, we find that using a
 larger pixel 
leads to a more efficient logarithmic mapping.
We compute the power spectra of the $3\times 1000$ convergence fields assuming periodic boundary conditions before and after logarithmic mapping. The covariance matrix is derived by calculating 
covariance 
between band powers at different 
wavenumber 
bins and at different 
source redshifts. The resulting covariance matrix represents dispersions
in the lensing power spectra for an area of $5^\circ \times 5^\circ$.
We assume a future, wide-field weak lensing survey of 5000 square
degrees, just as in the Dark Energy Survey~\citep[DES;][]{Abbott:2005bi},  
by rescaling each elements of the covariance matrix by $1/(5000/25)$.

\subsection{$l$ bin width}
For the power spectrum and the covariance calculations, we need to
determine the multipole bin width. An important requirement is that the
dimension of the covariance matrix is smaller than the total number of
realizations that are used for generating it.  The
dimension of the covariance matrix with three source redshift bins is
three times bigger than that with a single source redshift bin.  We
choose $\dl=200$ so that going up to $\lmax=2000$ requires a $30$ by $30$
covariance matrix. The dimension of the covariance matrix is then smaller than the 1000 realizations. 
Another reason for using $\dl=200$ rather than a smaller bin width is to reduce the sample variance effect in the derivatives calculation and therefore to make the derivatives smoother.

\subsection{Log-mapping}
We use the logarithmic mapping that was introduced in \paone. A log field is defined as:
\begin{equation}
\knew(\vec\theta) \equiv {\kappa_0}\ln\left[ 1 + \frac{\kappa(\vec\theta)}{\kappa_0}\right],\label{eq:knew}
\end{equation}
where $\kappa_0$ is a constant with a value slightly larger than the
absolute value of the minimum value of $\kappa$ for a given cosmology, source redshift,
and survey pixel -- this  keeps the argument of the logarithm positive.
In detail, to generate the covariance matrix for
the fiducial cosmology, we use $\kappa_0=|\kamin |+0.001$ where $\kamin$ is the minimum pixel $\kappa$ of the 1000 convergence fields at each source redshift: $\kamin=-0.01348$, $-0.02940$,
 $-0.04874$ at source redshifts $z_s=0.6$, 1, and 1.5, 
respectively.
While $\knew$ reduces to the standard convergence in the limit of small $\kappa$, the log alters $\kappa$ in very high or low convergence regimes.
The parameter $\kappa_0$ tunes the degree of the alteration 
such that, the smaller 
$\kappa_0$, the more we alter the field. Note that various properties of the 
$\knew$ field that will be discussed in this paper show a slow, asymptotic 
behavior as a function of $\kappa_0$ near the minimum value of $\kappa$. 
Therefore fine-tuning is not necessary for $\kappa_0$. 

\subsection{Derivatives}
For the Fisher matrix calculation, we need to first calculate numerical
derivatives of the lensing power spectra with respect to cosmological
parameters. To do this, we ran ray-tracing simulations of the
convergence fields for cosmological models perturbed around the fiducial
cosmology. 
We varied each of the following cosmological parameters: 
$As$,
 $n_s$, the cold dark matter density $\Ochh$ (the fiducial values for
$\Ochh=0.1054$), $\Ox$, and $w_0$ by $\pm 10\%$, respectively, 
and $w_a$ by $\pm 0.5$,
respectively. 
The Hubble parameter $h$, $\Om$, and $\Ob$\footnote{We hold $\Obhh$ fixed.} are then dependent parameters. For each of the 12 different cosmological models,  we built 40 realizations of the convergence fields for each of the three redshift bins. Therefore we use a total of $3(40\times12)=1440$ 
simulated convergence fields for calculating the derivatives.   We then computed
power spectra with and without the logarithmic transformation 
from each realization, averaged them, and derived the derivatives by differencing them. 
The value $\kamin$ (Eq.~\ref{eq:knew}) is derived for each cosmology and
therefore different for different cosmological models. 
Meanwhile, calculating derivatives becomes tricky in the presence of shape noise for the logarithmic transformation, as will be discussed in the next section.

\subsection{Shape noise}
The intrinsic ellipticities of source galaxy shapes cause white noise
contamination to the lensing power spectrum \citep{Kaiser:1996tp}. For the simulated
convergence map, we can include the noise contamination by adding, to
each pixel, random Gaussian noise with variance 
\begin{eqnarray}
\sigma_N^2=\frac{\sigma_\epsilon^2}{\bar{n}_g\Omega_{\rm pix}}, 
\end{eqnarray}
where $\sigma_\epsilon$ is the rms of intrinsic shear per
component, $\bar{n}_g$ is the mean number density of galaxies and
$\Omega_{\rm pix}$ is the pixel area. We set $\sigma_\epsilon=0.22$
and $\bar{n}_g=30~$arcmin$^{-2}$ as for our fiducial values 
at each redshift, therefore a total of 90 galaxies per arcmin$^{2}$
, which is much deeper than current weak lensing surveys such as Dark Energy Survey~\citep{Abbott:2005bi}. 
In the presence of shape noise, the fiducial, $\kappa$, field is closer to a Gaussian field and $|\kamin |$ becomes 
larger than without shape noise; the log-transform becomes less efficient. We measure $\kamin$ in the presence of the shape noise and use it when calculating the covariance matrix of the log-transformed field.

For the fiducial field before log mapping, we calculate the
derivatives from the power spectra without shape noise, because the
shape noise does not depend on cosmological parameters. That is, the
shape noise is included only in the covariance matrix. This is
consistent with what we would do in the likelihood analysis with real
data. We would derive a signal power spectrum after subtracting the
shape noise contribution \citep{Hikage:2010sq} 
and compare it with a theory power spectrum as
a function of cosmology.
 
Including the shape noise effect in the
derivatives calculation is non-trivial for the log-transformed
field. The shape noise not only increases noise in the measured
convergence field, but also makes the log-transform less efficient by
increasing $|\kamin |$. That is, when a convergence field is optimally
log-transformed by using $\kamin$, the power spectrum is different for a
different level of shape noise even after the shape noise is, ideally, subtracted.
In addition, while we conduct a nominal subtraction of the shape
noise from the power spectrum by subtracting a constant 
number that is derived from the difference in the variance of the field before and after the log-transform in the presence of shape noise, in truth, the effect of the shape noise is neither scale-independent nor cosmology-independent in the log-transformed field due to the nonlinear transformation. 
Therefore we need to derive the derivatives differently for different cosmologies for the log-transformed field. 
For each of the 12 cosmological model, we first estimated the 
$\kamin$ value for the log-transformed field with the shape noise. 
In order to minimize the impact of the random noise on the derivative
calculation, for each cosmology, we cloned each of the 40 realizations 
25 times using a different random seed for the shape noise. This way we generate 1000 realizations from the underlying 40 realizations for each cosmology; the averaged power spectrum will have much smaller sample variance on the effect of shape noise. We then log-transform the fields and differentiate the resulting power spectra between different
cosmologies and calculated the derivatives with respect to cosmological
parameters. 
 
\subsection{Fisher matrix}
We want to propagate the errors on the convergence power spectrum, our
observable, into the projections of cosmological parameters 
using a Fisher information matrix formalism. 
Using the numerical nonlinear derivatives and the
measured covariance matrices, 
we can compute the Fisher 
information matrix for cosmological parameters of
our interest, $p_i$, 
\citep{Tegmark1997,Tegmark1997etal}:
\begin{equation}
F_{ij}^{\rm WL}
\equiv 
\sum_{z_s,z'_s} \sum_{l,l'<\lmax}  \pderiv{\Cl l
	       {z_{s}}}{p_i}  \cov^{-1}(l,z_{s},l',z'_{s}) \pderiv{\Cl {l'}
	       {z'_{s}}}{p_j},
\end{equation}
where $\Cl l {z_{s}}$ is the measured convergence power spectrum at
multipole bin $l$ and $z_s$; \cov\ is the measured covariance matrix 
between wavenumbers and source redshifts. The set of cosmological parameters is: the amplitude of the power spectrum ($\ln A_S$), the slope of the primordial spectrum
($n_s$), the physical matter density in units of the critical density ($\Omhh$),  
the dark energy density $\Ox$, and two parameters for the dark energy equation of state: $w=w_0 + z/(1+z)w_a$.  Note that, by incorporating the
information from the three source redshifts into the Fisher matrix, 
we are including the tomographic lensing information \citep{Hu:1999ek,Takada:2003ef}.
The error on the $i$-th parameter including marginalization over
uncertainties in other parameters is estimated as
$\sigma(p_i)=\sqrt{
(\mbox{\boldmath $F^{-1}$})_{ii}}$,
where {\boldmath$F^{-1}$} is the inverse of
the Fisher matrix. The unmarginalized error is given as
$\sigma(p_i)=1/\sqrt{F_{ii}}$. 
We have also tried including independent, free shape noise as parameters for the three redshift bins, but find little effect on the constraints we derive.

\subsection{Planck prior}
Lensing information alone cannot determine all the cosmological
parameters simultaneously due to severe parameter degeneracies.  We
therefore include in these projections the CMB information expected from the Planck experiment.
The CMB is sensitive to two additional 
parameters, the baryon density $\Obhh$ and curvature density $\Ok$. We combine the convergence field information 
with the Planck prior 
\citep{Hu02}
by adding the $8\times 8$ Planck Fisher matrix to
the $6\times 6$ dimensional convergence Fisher matrix:
\begin{equation}
F_{ij} = F^{\rm WL}_{ij} + F^{\rm cmb}_{ij},
\end{equation} 
where the lensing Fisher matrix $F^{\rm WL}_{ij}$ 
has non-zero entries in the $6\times6$ block associated with $\ln As$,
 $n_s$, $\Ochh$ (or $\Oh$), $\Ox$, $w_0$, and $w_a$,
 and the Planck Fisher matrix $F^{\rm cmb}_{ij}$ has non-zero entries in all $8\times8$ elements 
including additional two parameters: $\Obhh$ and $\Ok$.
The Planck Fisher matrix we use includes
marginalization over uncertainties in the optical depth parameter $\tau$
on which the CMB power spectra depend. 
The zero $\Ok$ and $\Obhh$ 
elements in the convergence Fisher matrix reflect the fact that 
these two parameters are constrained solely by the CMB information. 

\begin{figure*}
\begin{center}
\includegraphics[width=6in]{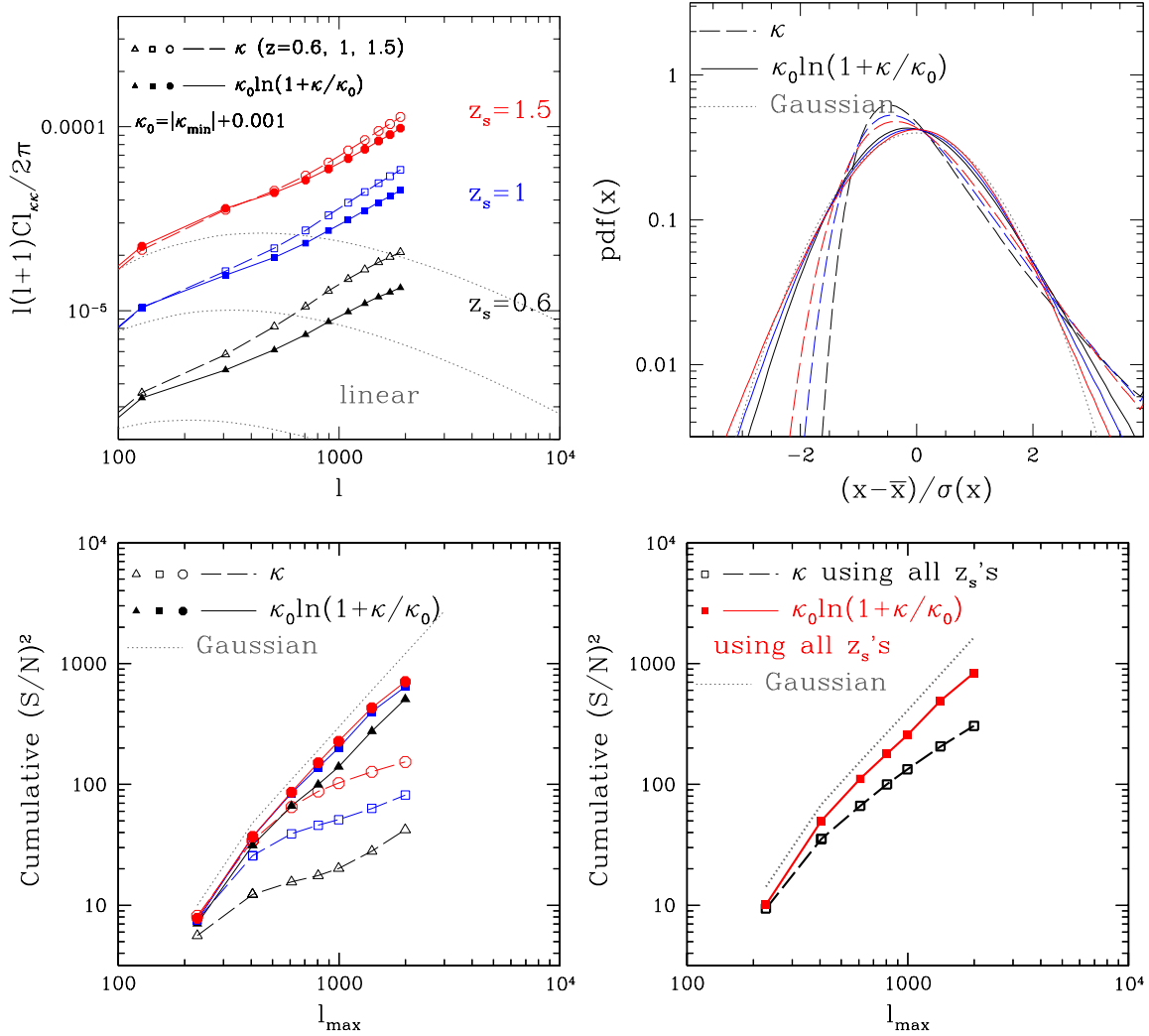}
\end{center}
\caption{Upper left: power spectra at $z=0.6$ (black), 1 (blue), and 1.5
 (red) before (dashed lines/open points) and after the log-transform (solid lines/filled points). Upper right: PDF distribution before and after the log-transform. Lower left: the information content of the individual redshift bins. Lower Right: the information content from the combination of the three $z_s$ bins. Since the three bins share information, the combination of the three bins (solid and dashed lines with points) show less improvement after the logarithmic transformation. Dotted lines show the Gaussian limit.  }\label{fig:IC}
\end{figure*}

\begin{figure*}[ht]
\begin{center}
\includegraphics[width=6in]{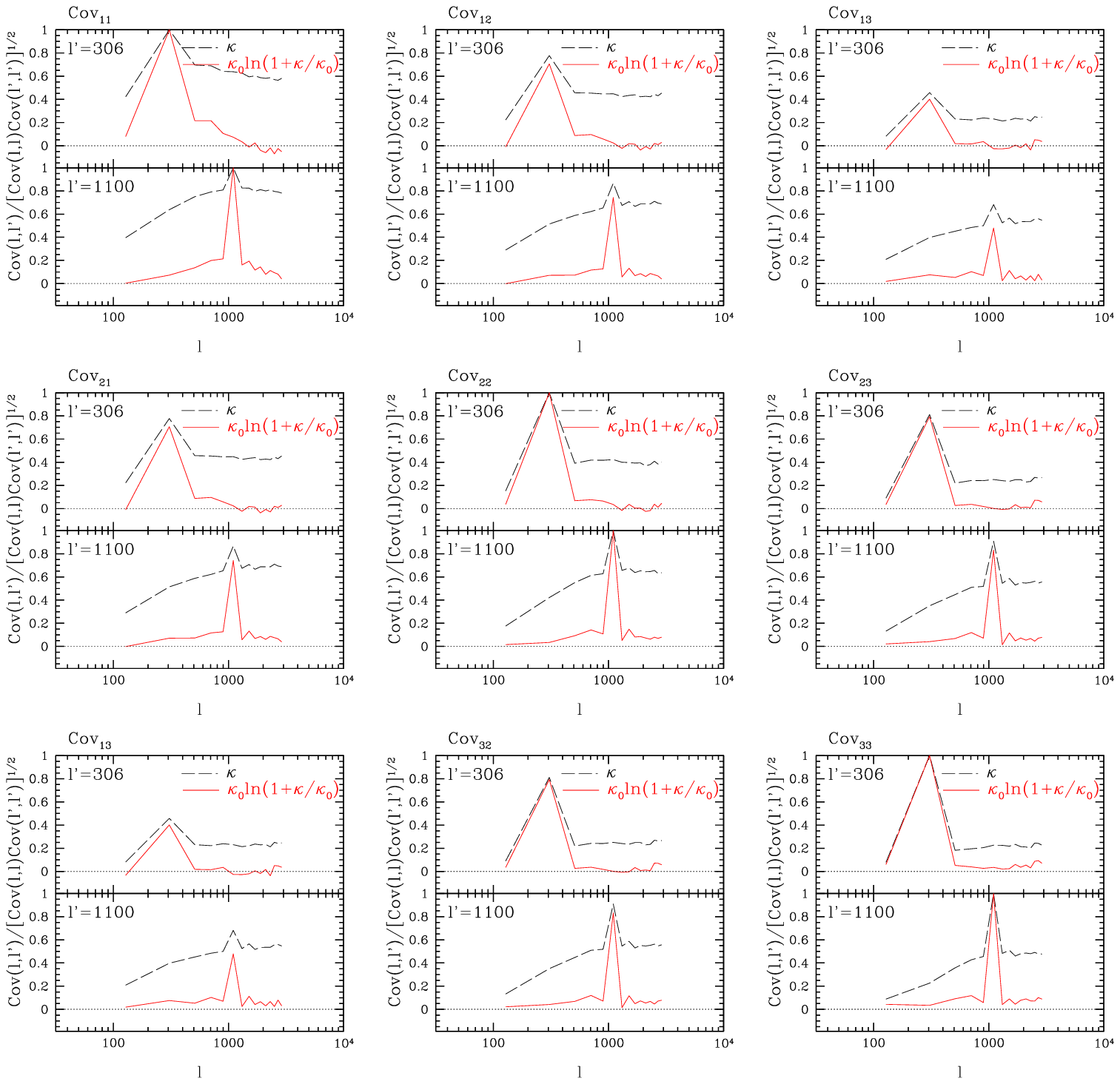}
\end{center}
\caption{Slices of the covariance matrix to show the covariance matrix property
 before (black dashed) and after (red solid) the log transform. Each
 panel shows a different block of the covariance matrix constructed by pairing two of the three $z_s$ 
bins. The elements are normalized relative to the diagonal elements, \ie, $\cov (l,l')/(\cov(l,l)\cov(l',l'))^{1/2}$. In the notation ${\rm Cov}_{ij}$, $i$ and 
$j$ indicates a source redshift bin. For example, ${\rm Cov_{11}}$ is the 
normalized auto covariance matrix of $z_s=0.6$. For each block of the 
covariance matrix, we find that the $\knew$ field is much more diagonal than 
$\kappa$. That is, the covariance between different $l$ bins is significantly 
reduced after the log-transform, while the covariance between different $z_s$ 
given $l$ remains similar before and after the log-transform.   }\label{fig:Cov}
\end{figure*}

\section{General properties of the log transformation}\label{sec:IC}
Before presenting the Fisher matrix analysis, we first revisit the 
general properties of the log-transform that have been discussed in \paone, 
such as the power spectra, 1-point probability distribution (PDF), the information content, and the structure of the covariance 
matrix for the three source redshift bins. 
 
The upper left panel of Figure \ref{fig:IC} shows the measured power spectra of the convergence field $\kappa$ (open symbols and/or dashed lines) and the log-transformed field $\knew$ (solid symbols and/or solid lines)
at $z_s=0.6$ (triangles and/or black), 1 (squares and/or blue), and 1.5 (circles and/or red). As expected, the log-transform reduces the small-scale nonlinear clustering. 
The reduction is more prominent for the lower source redshift, 
the reason for which is apparent in the 
PDF distribution in the upper right panel: the PDF of $\kappa$ deviates more from a Gaussian PDF at $z_s=0.6$ and therefore is more improved by the 
log-mapping. The lower left panel shows the improvement in the information 
content, 
the cumulative signal-to-noise ratio $(S/N)^2$ integrated up to a given
maximum multipole $l_{\rm max}$, for each of the three $z_s$ bins,
 defined as:
\begin{equation}
\left[\frac{S}{N}(\lmax)\right]^2 \equiv \left[ \sum_{l,l'<\lmax} C_l \cov^{-1}(l,l') C_{l'} \right]\label{eq:StoN}
\end{equation}
where $C_l$ is the power spectrum of multipole $l$  before and after the log-transform, $\cov$ is the covariance matrix describing correlations between 
the power spectra of multipoles $l$ and $l'$ ($l,l' < \lmax$) at each $z_s$,
 and the summation runs over all the
multipoles $l$ and $l'$ subject to $l,l'<\lmax$ \citep{Sato09,Takahashi09}.
As pointed out in \paone, this information content can be understood as the inverse of the fractional error on the amplitude of the observed, nonlinear power spectrum before and after the logarithmic transform. 
The Fisher matrix analysis will show to what extent this improved fractional
error on the observed amplitude remains when a full set of cosmological parameters is used. 
In the lower left panel, the improvement in the information content due to the log-transform is 
largest for $z_s=0.6$ and smallest for $z_s=1.5$ due to the level of
nonlinearities, 
as expected from the upper two panels.
The dotted gray line is the $(S/N)^2$ expected for a Gaussian case.
Due to the nonlinear structure growth 
 that causes significant off-diagonal covariances,  
 the measured $(S/N)^2$ values of the $\kappa$ field are much smaller than the Gaussian limit. We find that the $\knew$ field 
returns the $(S/N)^2$ closer to that of the Gaussian case, which confirms the
 results in \paone\ but using a different $\Delta l$. The improvement is a factor of $\sim 6.9$, 3.9, and 2.2 at $z_s=0.6$, 1, and 1.5, respectively, at $\lmax \sim 1000$ and a factor of 12, 7.9, 4.6, respectively at $\lmax \sim 2000$; note that the improvement is largest for the lowest source redshift. 
The improvement is slightly better than was reported in \paone, which is mainly due to the logarithmic transformation seemingly being more efficient for the larger pixels used here.
The lower right panel shows the improvement in the information content when 
the information from the three $z_s$ bins is combined. This is done by 
including the measured covariance between different $z_s$ bins in $\cov$ in 
Eq.~(\ref{eq:StoN}) and summing the signal-to-noise ratios up to $\lmax$ and over 
all the three $z_s$ bins. Since the three $z_s$ bins share
some of the lensing structures, 
there are non-vanishing covariances between the power spectra of
different $z_s$ bins and therefore the
 improvement due to the
log-transform is somewhat smaller when all $z_s$ are 
combined: 1.9 at $\lmax=1000$ and 2.7 at $\lmax=2000$. The Gaussian case in the lower right panel
 is derived also taking into account the expected covariance between different redshift bins.

\begin{figure*}[ht]
\begin{center}
\includegraphics{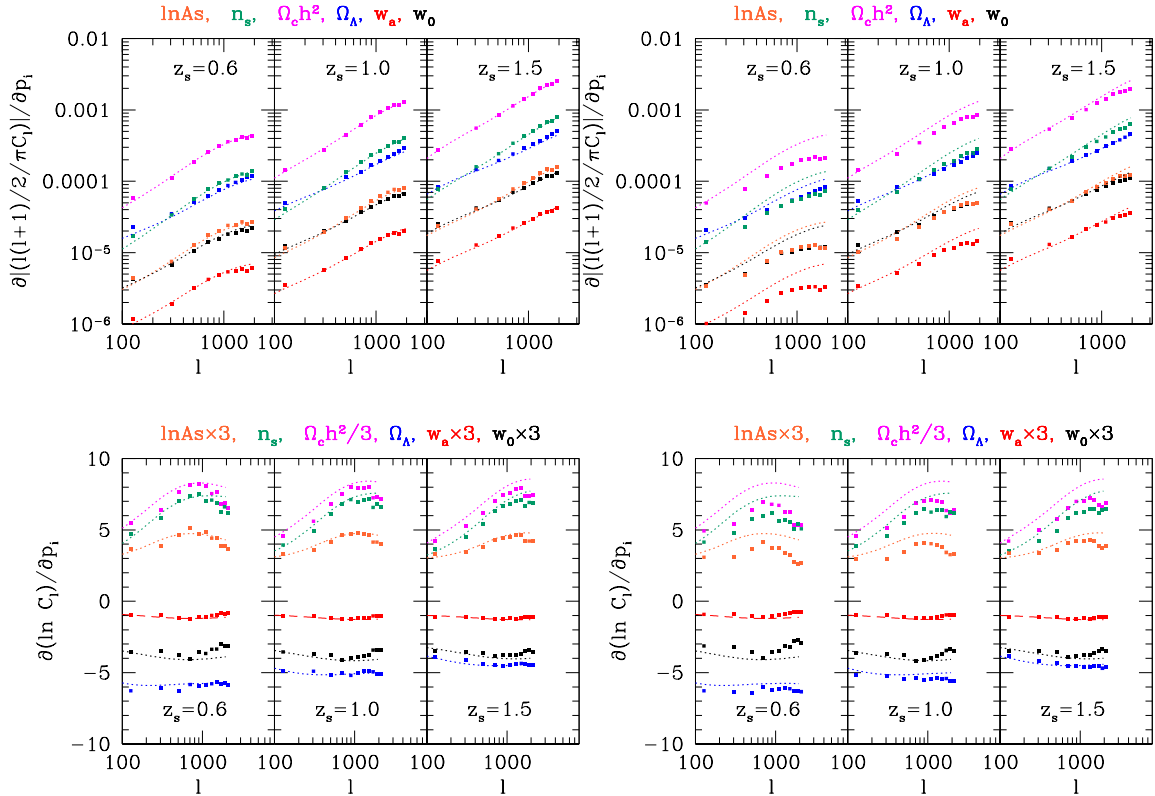}
\end{center}
\caption{Derivatives of convergence power spectra of $\kappa$ and $\knew$ fields with respect to various cosmological parameters in absolute values. Points are derivatives estimated from \Nb\ simulations; dotted lines use Halofit (for the fiducial map) as a comparison.
Left: the fiducial map. Right: the log-mapping. The fractional derivatives (bottom panels) remain similar even after the log-transform for some parameters, especially at high source redshifts. The fractional derivatives are rescaled by the factor denoted in the legends for clarity. 
}\label{fig:Deriv}
\end{figure*}

In addition to the improved information content, one of the potentially 
advantageous features of the log-mapping is the improvement in the covariance 
matrix property: it reduces the size of the off-diagonal terms (\paone). Figure 
\ref{fig:Cov} shows two rows of the covariance matrix for $l=306$ and $l=1100$. 
Each panel shows a different 
block of the normalized covariance matrix constructed by pairing two of the 
three $z_s$ bins. For each block of the covariance matrix, we find that the 
$\knew$ field is much more diagonal than $\kappa$. That is, the covariance
 between different $l$ bins is significantly reduced by the log-transform.
Both $\kappa$ and $\knew$ fields show a slightly higher level of 
off-diagonal covariance compared to what we 
found in \paone, partly due to the larger $\dl$ bin used here (i.e., $\dl=200$ compared to $\dl=100$ in \paone) 
and also probably due to sample variance. 
Note that the different bin width alters only the Gaussian covariance
contribution \citep{Meiksin99,Sco99,Cooray01,Takada09}; the larger bin width reduces
the Gaussian covariance, the diagonal components of the covariance
matrix, 
and thus increases the
relative off-diagonal components. 
While the covariance between different wavenumbers is
decreased, the 
covariance between different $z_s$ given $l$ remains similar before and 
after the log-transform by looking at the location of the peaks in ${\rm Cov}_{ij}$ for $i \ne j$.

In summary, we observe general properties of the log-transform that are consistent with the results in \paone: the 1-point PDF is more Gaussian, the covariance matrix is closer to a diagonal matrix, and the information content is greatly improved after the log-transform. We next propagate this improvement to the errors on cosmological parameters using the Fisher matrix formalism.

\begin{figure}[ht]
\begin{center}
\includegraphics[width=3in]{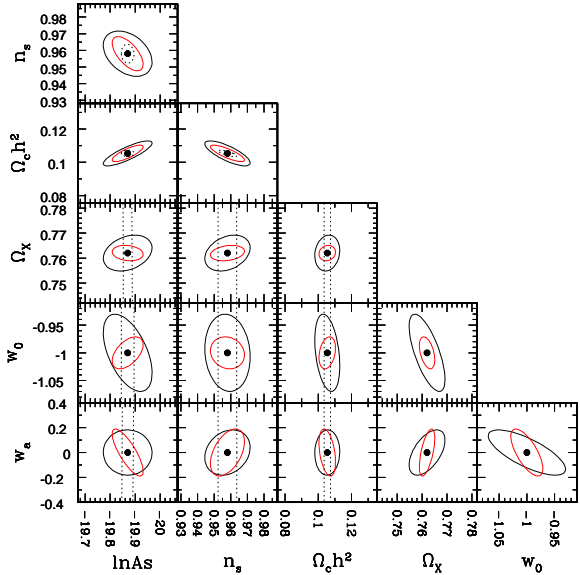}
\end{center}
\caption{Marginalized error contours before (solid black lines) and after
 log-transform (red lines) without shape noise contamination and CMB information.
That is, we took the $2\times2$ sub-Covariance matrix and plotted the error contours.
We use power spectrum information up to $l=2000$. One sees that 
the marginalized errors are overall smaller for $\knew$. 
The dotted contours show constraints from the
 Plank alone. The Planck contours for pairs between $\Ox$, $w_0$, and $w_a$ do not show up here due to the extreme degeneracies.
}\label{fig:NoPlanck}
\end{figure}

\begin{deluxetable*}{c|cccccccc}
\tablewidth{0pt}
\tabletypesize{\footnotesize}
\tablecaption{\label{tab:tabmarg} Marginalized errors on each parameters.}
\startdata \hline\hline
&$\lnAs$  & $n_s$   & $\Omhh$  &$\Ox$   & $w_0$   &  $w_a$  & $\Obhh$ &  $\Ok$ \\\hline
$\kappa$ alone & 0.0642 & 0.897E-02 & 0.493E-02 &  0.472E-02 &  0.458E-01 &  0.119  \\
$\knew$  alone & 0.0408 &  0.674E-02 &  0.325E-02 & 0.202E-02 & 0.189E-01  & 0.124  \\ \hline
$\kappa$ + Planck& 0.0101 &  0.336E-02  & 0.115E-02 &  0.367E-02 &  0.357E-01 & 0.997E-01 & 0.150E-03 & 0.272E-02 \\
$\knew$ + Planck& 0.0102  & 0.304E-02  & 0.107E-02 & 0.149E-02 & 0.161E-01 &  0.632E-01 & 0.148E-03 & 0.253E-02
\enddata
\tablecomments{Marginalized errors before and after the log-transform
 using power spectrum information up to $\lmax=2000$. 
Both errors show distinct improvement after log-transform; the improvement is mainly on $\Ox$, $w_0$ and $w_a$ once the Planck prior is included.}
\end{deluxetable*}

\begin{figure}[ht]
\begin{center}
\includegraphics[width=3.5in]{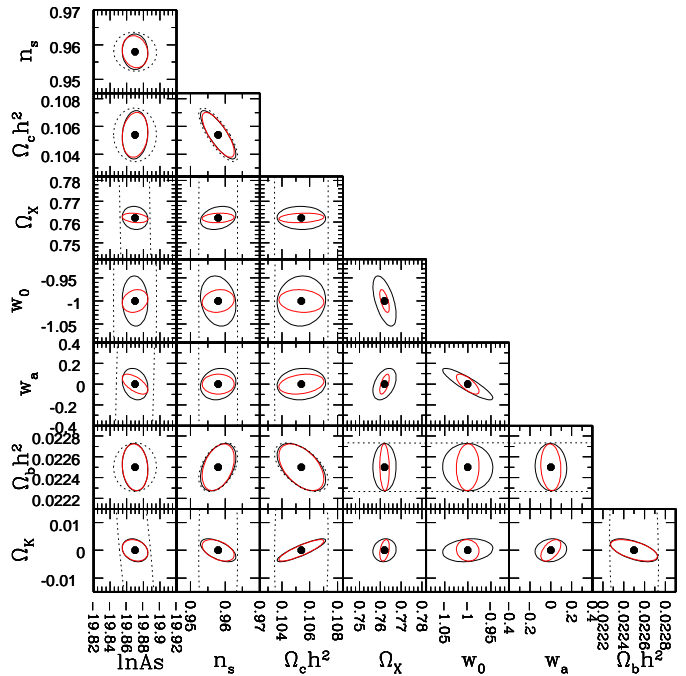}
\end{center}
\caption{Marginalized error contours before (solid black lines) and after
 log-transform (red lines) when we combine the
 weak lensing data with the Planck mission. We use power spectrum information up to $l=2000$. One sees that $\knew$ improves constraints mainly on the three dark energy parameters, i.e., $\Ox$, $w_0$, and $w_a$.}\label{fig:withPlanck} 
\end{figure}

\section{Fisher matrix analysis}\label{sec:Fisher}
 We study how the improvement on the precision of the amplitude, 
or the information content, propagates into
the precision of cosmological parameters.
The improvement in the information content for the $\knew$ field is due to the improved properties 
of its covariance matrix.  
The Fisher matrix formalism then combines this with an extra piece: the sensitivity 
  of the $\kappa$ and $\knew$ power spectra to cosmological parameters.

Before presenting the Fisher 
matrix results, we take a look at the numerical derivatives calculated from 
the convergence fields from a large set of \Nb\ simulations.
Figure \ref{fig:Deriv} shows the derivatives of the power spectrum of $\kappa$ (left panels) and $\knew$ (right panels) fields with respect to various cosmological parameters (square points). As a comparison, the dotted lines show the prediction based on Halofit \citep{Smith03} for the fiducial mapping. We see an obvious decrease in the relative amplitude of the derivatives due to the log-transform (bottom panels) at $z_s=0.6$. Such changes in the derivatives will be combined with the changes in the covariance matrix property in the Fisher matrix calculation. As a caveat, we find the relative amplitude of the derivative with respect to $n_s$ decreases after the log-transform, while \citet{Neyrinck11c} finds an increase on small scales; it might be due to a difference in details of the log-transform between the weak lensing field and the density field.

\begin{figure*}[ht]
\begin{center}
\includegraphics[width=6in]{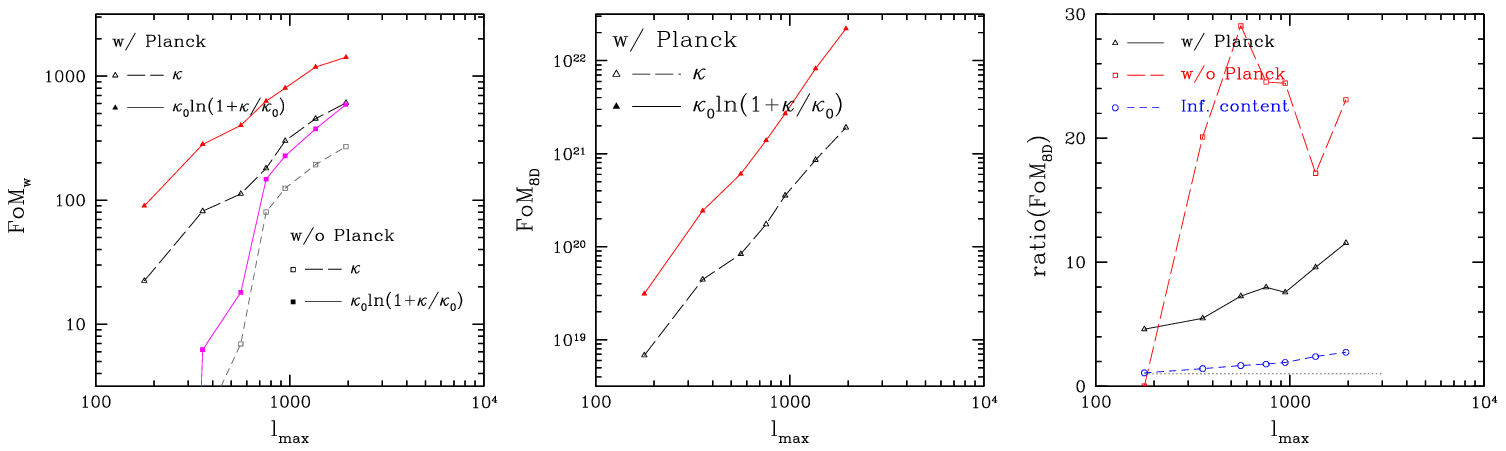}
\end{center}
\caption{Figure of Merit in $w_0$-$w_a$ (Left) and in the 8-D (middle) before and after the log-transform. The magenta and gray lines with squares in the left panel show the result without the Planck prior while the red and black triangles show the values with the Plank prior. We find more than a factor of 2 improvement in $\FoMw$.
The middle panel shows a 8-dimensional figure-of-merit, i.e., the inverse volume of the 8-dimensional parameter space. The right panel shows the ratio of the 8-D FoM between with and without the log-transform. We find a factor of 7-12 improvement using $l_{\rm max}=1000-2000$ in the presence of the Planck prior.}\label{fig:FoM}
\end{figure*}

\subsection{Without Planck prior}
We first show the Fisher matrix results of the convergence field without 
shape noise contamination and 
CMB information in Figure \ref{fig:NoPlanck}. We use all the
information up to $l=2000$. For reference, if the amplitude of the power spectrum were the only parameter (as in Paper I), the log field would lead to an error $\Delta \ln(A_s)=1.9\times 10^{-3}$ for these survey parameters while the standard $\kappa$ field would have $\Delta \ln(A_s)=2.9\times 10^{-3}$. We call this a factor of 1.5 improvement in the {\it 1D Figure of Merit}. When we generalize to 6 parameters, the corresponding Figure of Merit (FoM, hereafter) is the square root of the determinant of the $6\times6$ Fisher matrix. In this case, we find a factor of 23 improvement. This is better than the naive expectation of $1.5^6=11$, so our first conclusion is that the advantages of the log estimator hold up -- or even increase -- when generalizing to multiple cosmological parameters.

Figure \rf{NoPlanck} shows some 2D slices of these constraints.
The figure shows the
marginalized $1-\sigma$ error contours from weak lensing tomography
alone for various pairs of the 6 cosmological parameters\footnote{I.e.,
we take the 
$2\times2$
sub-Covariance matrix and plot the error contours.}. The solid black lines show the result from the $\kappa$ field before the log mapping and the red lines show the results of the $\knew$ field. The constraints are tighter for the $\knew$ field: the error ellipses have shrunk and the projections are that the $\knew$ field often leads to narrower allowed regions. 
Table \ref{tab:tabmarg} lists the marginalized errors without the Planck prior: the log-transform shows improvement, especially on $\Ox$ and $w_0$.  The dotted contours in Figure \ref{fig:NoPlanck} show constraints from Planck alone: the Planck contours for pairs between $\Ox$, $w_0$, and $w_a$ do not show up here due to the extreme degeneracies. For all parameters other than the dark energy parameters, the information from the Planck mission dominates. 
 
\subsection{With Planck prior}
The situation is qualitatively similar when the Planck prior is added in. The Figure of Merit (now 8D) is larger by a factor of 12 when the log estimator is used. 
Figure \ref{fig:withPlanck} shows slices of the
 error contours on the 8 cosmological parameters when we combine the
 weak lensing data with Planck (left panel). 
The solid black line is for the $\kappa$ field and the red line for the $\knew$
 field after the Planck Fisher matrix is combined. The dotted contour
 shows the error contours for the Planck alone as a comparison. When the
 Planck priors are included, the information on parameters other than $\Ox$, $w_0$, and $w_a$ is
 dominated by the Planck information, as evident in Figure
 \ref{fig:NoPlanck}; $\knew$ improves constraints mainly on the three
 dark energy parameters~\footnote{The solid contours in Figure
 \ref{fig:withPlanck} do not exactly agree with the naive error contour
 combination of the solid contours in Figure \ref{fig:NoPlanck} and
 the dotted Planck contour: we would expect this agreement if we add the
 Fisher matrices from the 2 by 2 sub-covariance matrices of the
 convergence field and the Planck mission. However we are adding the two
 full Fisher matrices here such that a given marginalized error contour
 is affected by the effect of the Planck information on the rest of the
 parameters.}.
 Table \ref{tab:tabmarg} shows that we achieve an improvement by a factor of 2.5 for $\Ox$, 2.2 for $w_0$, and 1.6 for $w_a$ by log-transformation.

\begin{figure*}[ht]
\begin{center}
\includegraphics[width=6in]{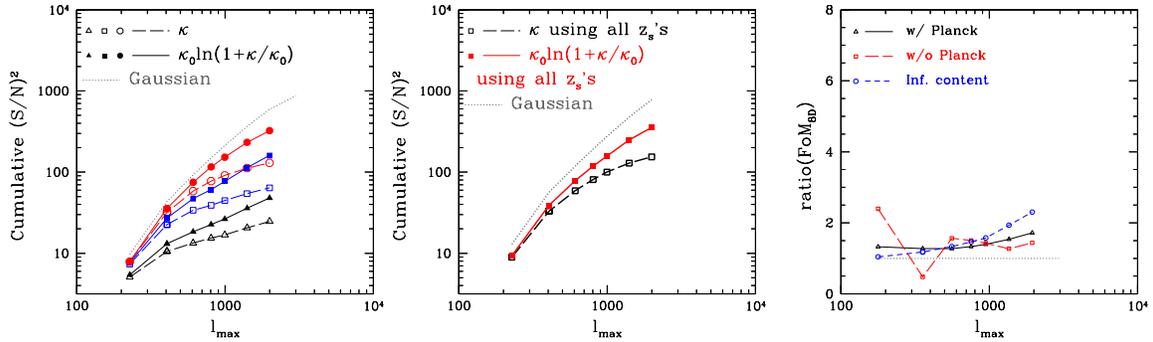}
\end{center}
\caption{Left: the information content in the presence of shape noise ($\bar{n}_g=30 \iams$ for each redshift bin). The dotted line shows the Gaussian limit in the presence of shape noise. Middle: the information content from the combination of the three source redshift bins. Right: the 8D FoM. We find very little improvement in the constraints on the final cosmological parameters once shape noise is added.}\label{fig:StoNshape}
\end{figure*}

\begin{figure}
\begin{center}
\includegraphics[width=3.5in]{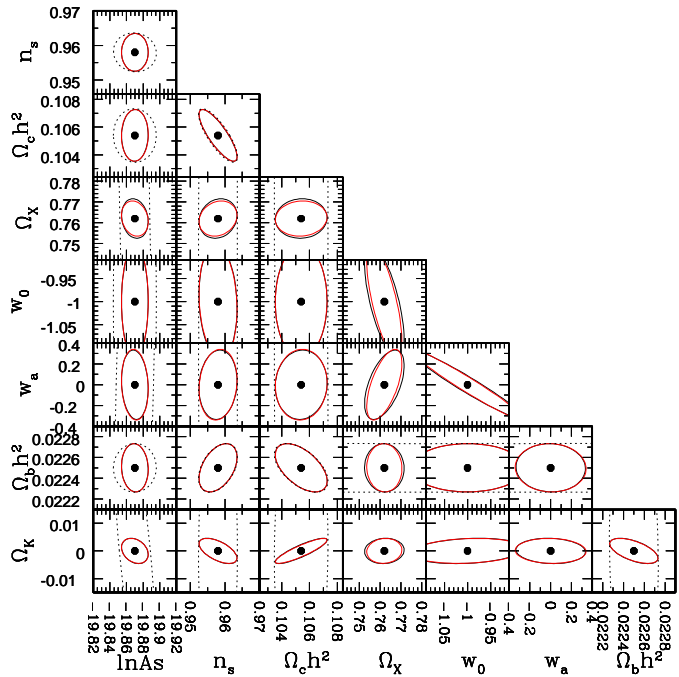}
\end{center}
\caption{In the presence of shape noise ($\bar{n}_g=30 \iams$ for each redshift bin), we find little improvement in the constraints on the
 final cosmological parameters after log-transform. Black: before log-transform. Red: after log-transform.}\label{fig:Errorshape}
\end{figure}

\subsection{Improvement in the dark energy FoM}
We can also quantify the improvement due to the log-transform using the 
Figure-of-Merit of dark energy parameters (hereafter `$\FoMw$')
that is often used in literature to characterize the performance
of a dark energy survey mission.
The Dark Energy Task Force (DETF) FoM
\citep{DETF} is defined as
\begin{equation}
{\rm \FoMw}\equiv \frac{1}{\sigma(w_p)\sigma(w_a)}=\frac{1}{
\sqrt{\det({\rm Cov}[w_0,w_a])}},\label{eq:FoM}
\end{equation}
where $w_p$ is the dark energy equation of state at the ``pivot''
redshift, at which the dark energy equation of state is best constrained
by given observables, and ${\rm Cov}[w_0,w_a]$ is the $2\times 2$
sub-matrix of the inverted Fisher matrix, $\mbox{\boldmath$F^{-1}$}$,
including only its elements of $w_0$ and $w_a$. The FoM is proportional to the area of
the marginalized error ellipse in $w_0$ and $w_a$ parameter sub-space. 
 In Figure \ref{fig:FoM}, the left panel shows $\FoMw$ before (dashed lines) and after the
log-transform (solid lines): 
we observe approximately a factor of two improvement when including the Planck prior (red and black lines with triangles).

In the middle panel, we show the 8-dimensional figure-of-merit as a function of $\lmax$. We find a factor of 7-12 improvement using $l_{\rm max}=1000-2000$, which can be read out from the right
panel where we show the ratio of ${\rm FoM}_{8D}$ between with and
without the log-transform (black line). Table \ref{tab:tabFoM} presents 
the FoM values
as a function of $l_{\rm max}=1000-2000$.
\begin{center}
\begin{deluxetable}{c|cc|cc}
\tablewidth{0pt}
\tabletypesize{\footnotesize}
\tablecaption{\label{tab:tabFoM} Figure of Merit.}
\startdata \hline\hline
$l_{\rm max}$ & $\FoMw$ &   & ${\rm FoM}_{8D}$ & \\
 & $\kappa$ & $\knew$ & $\kappa$ & $\knew$ \\ \hline
1000 & 301 & 803 & 3.56E+20 & 2.67E+21 \\
2000 & 610 & 1423 & 1.90E+21 & 2.21E+22 
\enddata
\tablecomments{Figure of Merit in $w_0$-$w_a$ and in the 8-D before and after the log-transform.}
\end{deluxetable}
\end{center}

\begin{figure*}[ht]
\begin{center}
\includegraphics{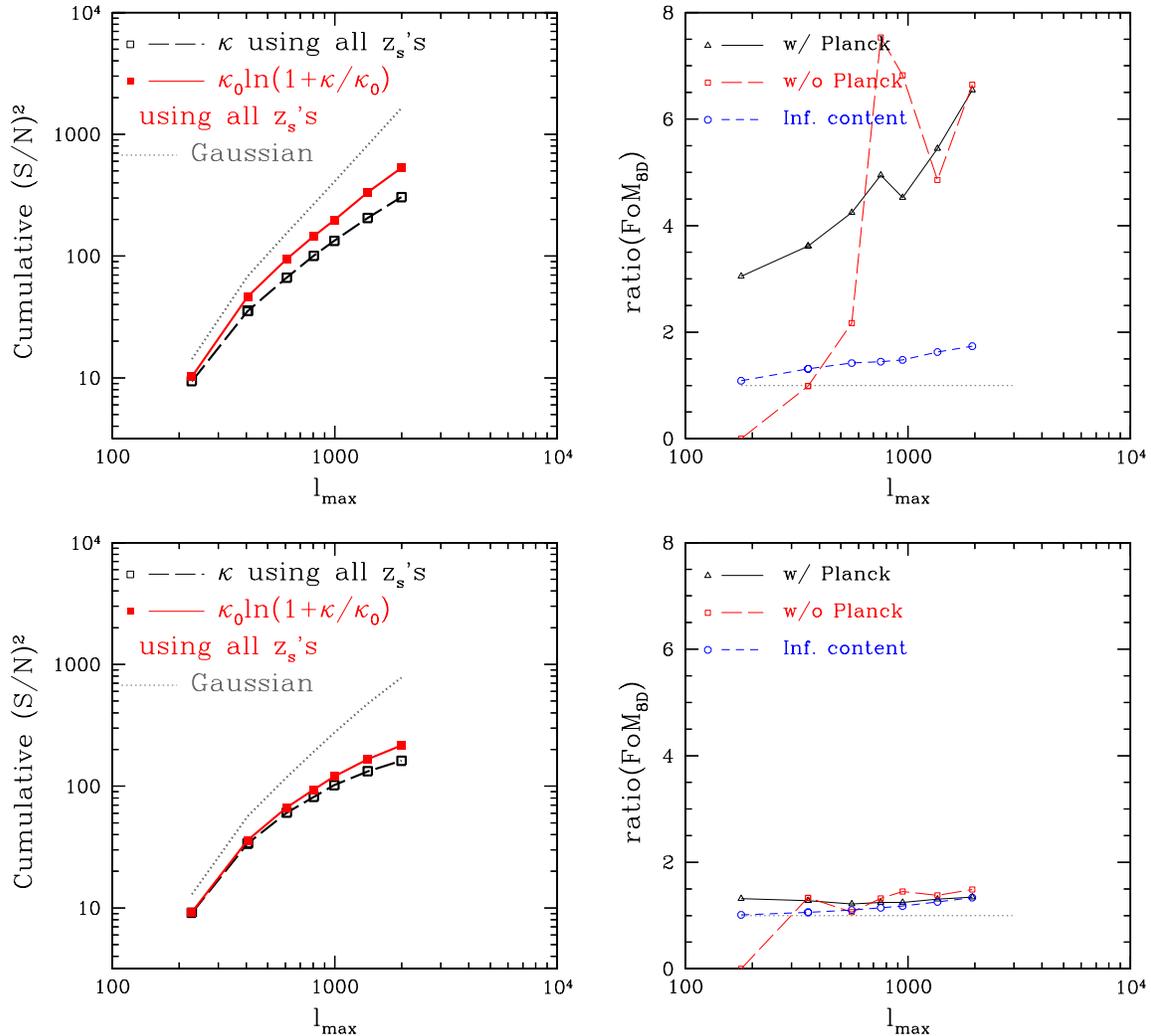}
\end{center}
\caption{Effects of pixel size. We use a pixel of 0.6 arcmin for this figure. Top: without shape noise. Bottom: with shape noise ($\bar{n}_g=30 \iams$ for each pixel). Left: the information content. Right: improvements due to log-mapping in terms of various quantities.}
\label{fig:StoNpixel}
\end{figure*}

\begin{figure*}[ht]
\begin{center}
\includegraphics[width=7in]{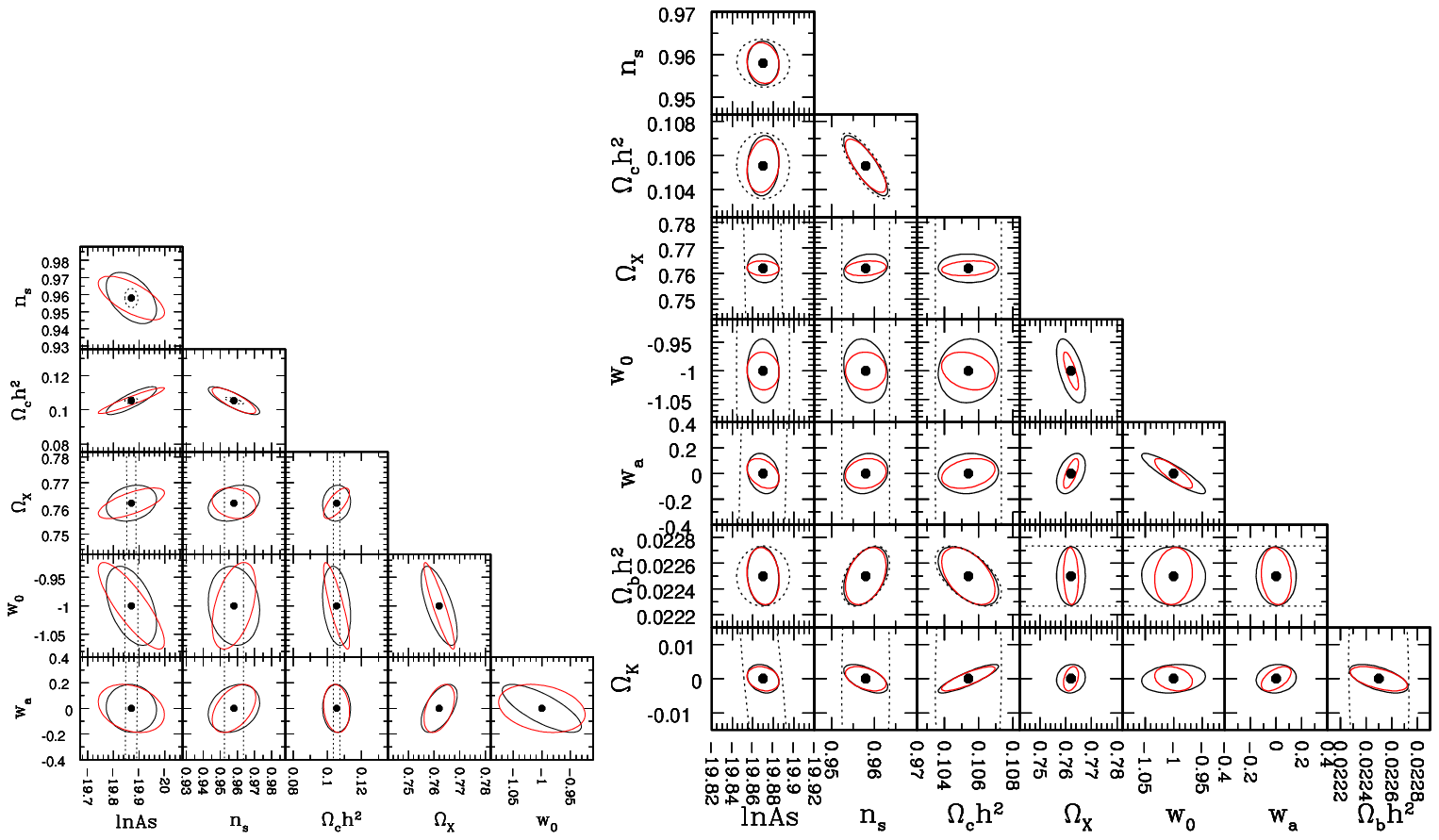}
\end{center}
\caption{Effects of pixel size. Using a pixel of 0.6 arcmin. Marginalized error contours before (solid black lines) and after
 log-transform (red lines) without shape noise contamination. Left: without the Planck prior. 
Right: with the Planck prior.
We use power spectrum information up to $l=2000$. One sees that the log-transform is less efficient with a smaller pixel.
}\label{fig:Errorpixel}
\end{figure*}

\section{Fisher matrix analysis with shape noise}\label{sec:Fshape}
\def \iams {{\rm arcmin}^{-2}}
\def \lmax {l_{\rm max}}
Any weak lensing survey contains shape noise,
i.e., uncertainty associated with 
intrinsic shapes of the galaxies. This noise decreases with
increasing galaxy number density. In the presence of large shape
noise, the observed field is closer to Gaussian (assuming Gaussian
shape noise) and $\kamin$ is larger due to the additional dispersion: we
therefore expect that a log-mapping will be less efficient for a larger shape
noise. \paone\ have shown that, with a galaxy number density of $30~
\iams$ at $z_s=1$ and a pixel size of 2.4 arcmin, a factor of the improvement is 2.4 for $\lmax=2000$; we find a similar result of 2.3 as shown in the left panel of Figure \ref{fig:StoNshape}. 
We assume the total mean number density of $90~ \iams$ and the number
densities of 30, 30, and $30~\iams$ for source redshifts 
of $z_s=0.6$, 1.0, and 1.5, respectively. Note that this is a galaxy number density that is much higher than ground-based, near-future weak lensing surveys.

Figure \ref{fig:StoNshape} shows that, in the presence of this modest shape noise, the improvement in the information content due to the log-mapping is still large, but the 8D FoM is very similar for the standard $\kappa$ estimator and for $\knew$. The improvement on the full set of cosmological parameters is only a factor of 1.7 for $\lmax=2000$.

Figure \ref{fig:Errorshape} shows the marginalized error contours with the Planck prior in the presence of shape noise: the improvement due to the log-mapping is not found here despite the large degree of improvement in the information content. 
Therefore, we find the improvement in the information content by 
the log-mapping does not propagate efficiently to the improved cosmological information in the presence of even an optimistic limit of shape noise for future weak lensing surveys. 
The effect of shape noise we find is consistent with \citet{Joachim11} despite the different set of cosmological parameters investigated, except that they find much smaller improvement in the information content. In calculating the information content (Eq. \ref{eq:StoN}), we set the signal to be an averaged power spectrum of the log-transformed field after subtracting a constant power as an approximation for the shape noise effect; \citet{Joachim11} use a power spectrum without shape noise as a signal. Since the constant power we subtract does not include higher order contributions that mingles shape noise and the clustering signal, the signal we input is higher than that of \citet{Joachim11} and therefore our information content is larger. 

It is quite possible that the problem of the shape noise lies in the estimator. In the absence of shape noise, the log-transformed field is the obvious way to make the field nearly Gaussian. In the presence of shape noise, it is quite possible that one must work harder to find an estimator that recaptures the information lost to higher point functions. A different Gaussianization scheme, particularly one not so sensitive to $\kamin$, might well work better. It is also possible that the information is hopelessly lost due to the shape noise and there is not much to recover with log-transform.
The encouraging results to date in the absence of shape noise suggest that this is an important avenue of research.

\section{The effect of the pixelization}\label{sec:Pixel}
Using a larger pixel appears more efficient in terms of the information content
than using a smaller pixel, whether with and without shape noise, partly due to the smaller $|\kamin|$ that we can reach for a larger pixel. 
The top panels of Figure \ref{fig:StoNpixel} shows the information content and the 8D FoM using a pixel of 0.6 arcmin, instead of our fiducial pixel of 2.4 arcmin, in the presence of no shape noise. From the left panel, one sees that the information content after log-mapping (left panels) is much lower when using a smaller pixel than when using a pixel of 2.4 arcmin (in Figure \ref{fig:IC})~\citep[also see][]{Neyrinck11b,Joachim11}.  On the other hand, we find that the difference in ${\rm FoM}_{8D}$ for different pixel sizes is not as drastic as we have expected based on the information content result. The left panel of Figure \ref{fig:Errorpixel} shows the marginalized error contours for the pixel of 0.6 arcmin without shape noise and without the Planck prior. Since this case is less efficient for the log-transform, unlike the case of a 2.4 arcmin pixel (in Figure \ref{fig:NoPlanck}), the changes in the error ellipses are not a simple shrinkage of ellipses. The projections are that the $\knew$ field leads to narrower allowed regions, which implies stronger degeneracies after the log-transform; the errors on cosmological parameters do not decrease after the log-transform despite the improvement in the information content.  When the Planck priors are combined (right), the strong degeneracies in the power spectrum of $\knew$ are lifted, and now the thin error ellipses observed in the left panel finally translate to smaller errors on $\Ox$, $w_0$, and $w_a$.

The bottom panels of Figure \ref{fig:StoNpixel} shows the
information content and the 8D FoM using a pixel of 0.6 arcmin {\it in
the presence of shape noise}. In comparison to Figure
\ref{fig:StoNshape} for  pixel of 2.4 arcmin, we find even less
improvement with the log-transform due to a smaller pixel. The effect of
pixel we observe appears consistent with the effect of smoothing in \citet{Joachim11}.

\section{Analytic Fisher matrix results}\label{sec:Fhalofit}
We compare our \Nb\ Fisher matrix results with the analytic Fisher matrix results using a Halofit \citep{Smith03} for the fiducial mapping. As shown in the left panel of Figure \ref{fig:Deriv}, the derivatives derived from the Halofit slightly deviates from the \Nb\ results, especially on high $l$, such that those for the dark energy parameters appear similar to what was shown in~\citet{Casa11}. Due to the discrepancy in the nonlinear convergence power spectrum between the Halofit and the \Nb\ result~\citep{Eifler11}, there seems to be a bigger difference in the fractional derivatives (bottom left panel). We conduct a Gaussian Fisher matrix analysis using Halofit results and the Gaussian assumption: we call this `Gaussian Fisher matrix results'. 

We find the \Nb\ Fisher result predicts better constraints on dark
energy parameters than the Gaussian Fisher result, which is contrary to
\citet{Casa11}. 
In detail, the nonlinear covariance matrix increases the error bars,
relative to the Gaussian case, which is expected; however, the \Nb\
results appear to show less degeneracies between dark energy parameters ($\Ox,
w_0, w_a$) than predicted by
Halofit. 
As a result, the \Nb\ Fisher
matrix analysis gives better constraints on dark energy parameters than
the Halofit-based Gaussian case. 
If we hold $w_a$ fixed or $\Ox$ fixed, the \Nb\ FoM 
becomes worse than the Halofit result.
In the presence of shape noise, the nonlinear covariance matrix is
closer to the Gaussian one. Therefore the constraints from the \Nb\
result are better than the Halofit-based Gaussian case to a larger
extent. Table \ref{tab:halofit} lists our results. \begin{deluxetable*}{c|cccccccc}
\tablewidth{0pt}
\tabletypesize{\footnotesize}
\tablecaption{\label{tab:halofit} \Nb\ Fisher analysis vs analytic Gaussian Fisher analysis.}
\startdata \hline\hline
&$\Ox$ & $w_0$   &  $w_a$  \\\hline
$\kappa$ + Planck&  0.367E-02 &  0.357E-01 & 0.997E-01  \\
Analytic        & 0.372E-02 & 0.451E-01 &  0.128  \\
$\kappa$ + Planck & 0.335E-02  &  0.165E-01 & -- \\
Analytic        & 0.154E-02  & 0.154E-01 & -- \\
$\kappa$ + Planck &  --   &0.294E-01 & 0.909E-01 \\         
Analytic        & --         & 0.171E-01 & 0.530E-01 \\ \hline
$\kappa$ + Planck, shape noise&  0.626E-02 &  0.808E-01 & 0.220  \\
Analytic, shape noise        & 0.134E-01 & 0.169 &  0.409  \\
$\kappa$ + Planck, shape noise & 0.525E-02  &  0.211E-01 & -- \\
Analytic, shape noise        & 0.429E-02  & 0.237E-01 & -- \\
$\kappa$ + Planck, shape noise &  --   &0.603E-01 & 0.185 \\         
Analytic, shape noise        & --         & 0.409E-01 & 0.131 \\ \hline
\enddata
\tablecomments{The analytic Gaussian Fisher results are derived using the Halofit results and the Gaussian assumption. Shape noise is based on $\bar{n}_g=30~\iams$ for each source redshift bin. The results on parameters other than dark energy parameters are very similar between the two methods, mainly because these are dominated by the Planck information; we do not show these parameters in this table for simplicity. The rows with ``--'' means that the corresponding parameter is held fixed.}
\end{deluxetable*}

We comment that 
sample variance introduces a noisy feature in the \Nb\ derivatives. We have used $\dl=200$ rather than a smaller bin width to reduce such noisy feature. However, any remaining noise might have affected the \Nb\ results. We tried smoothing the derivatives, which did not decrease the constrains from the \Nb\ results and therefore did not reverse our finding. 

\section{Conclusion}\label{sec:con}
We have used the Fisher matrix formalism to test the impact of the log-transform on the cosmological parameters. We find that the log mapping performs much better than the fiducial mapping: when Planck mission information is included as a prior, the log-transformed field greatly improves constraints especially on dark energy parameters such as $\Ox$, $w_0$, and $w_a$. 
In the presence of shape noise, however, the advantage of the log
mapping quickly diminishes. We find little improvement on the
cosmological parameters after log-transform even with $\bar{n}_g=30 \iams$ at
each of the three source redshift bins. We find the information
content is not necessarily a good probe of the actual precision on the
final cosmological parameters. We also find using a larger pixel allows
a more efficient log-transform with and without shape noise. Finally, we
find that, for the fiducial mapping, the Halofit-based Gaussian Fisher
matrix calculation gives worse constraints on dark energy parameters
than the full \Nb\ Fisher matrix result. This appears to be due to 
less degeneracies among dark energy parameters in the \Nb\ power
spectra. 

\acknowledgements
We greatly appreciate the extremely helpful comments from Bhuvnesh Jain.
We thank Patrick McDonald and Jan M. Kratochvil for useful discussions.
H-JS is supported by the Berkeley
Center for Cosmological Physics. 
MS is supported by a Grant-in-Aid for JSPS fellows.
MT is supported by the Grants-in-Aid for Scientific
Research Fund (No. 23340061), JSPS Core-to-Core Program
``International Research Network for Dark Energy'', World Premier
International Research Center Initiative (WPI Initiative), MEXT, Japan,
and the FIRST program ``Subaru Measurements of Images and Redshifts
(SuMIRe)'', CSTP, Japan. 
SD is supported by the US Department of Energy, including grant DE-FG02-95ER40896; and by National Science Foundation Grant AST- 0908072.

\end{document}